\documentclass[a4paper,10pt]{article}

\usepackage{amsmath,amssymb,amsbsy,latexsym,amsthm}
\usepackage{graphicx}
\usepackage{psfrag}
\usepackage{setspace}
\usepackage[all]{xy}

\theoremstyle{plain}

\newcommand{\eqa}{\begin{eqnarray}}
\newcommand{\neqa}{\end{eqnarray}}
\newcommand{\be}{\begin{equation}}
\newcommand{\ee}{\end{equation}}

\newcommand{\bea}{\begin{eqnarray}}
\newcommand{\eea}{\end{eqnarray}}

\def\R{\mathbb{R}}

\def\C{\mathbb{C}}

\def\1{1 \!\! 1}

\def\ra{\rangle}
\def\la{\langle}

\newcommand{\mean}[1]{\la{#1}\ra}

\newcommand{\SU}{\mathrm{SU}}
\newcommand{\SO}{\mathrm{SO}}

\newcommand{\U}{\mathrm{U}}

\newcommand{\lalg}[1]{\mathfrak{#1}}
\newcommand{\su}{\lalg{su}}

\newcommand{\Ref}[1]{(\ref{#1})}

\newcommand{\hO}{\hat{O}}
\newcommand{\hP}{\hat{P}}
\newcommand{\hh}{\hat{h}}

\usepackage{dsfont}
\newcommand{\ie}{{\it i.e.}~}
\newcommand{\eg}{{\it e.g.}~}
\newcommand{\po}{{\mathds{P}}}

\newcommand{\id}{\mathbb{I}}
\newcommand{\cp}{{t}}

\usepackage{color}
\usepackage{graphicx}
\newcommand{\lp}{{\ell_{\mathrm{P}}}}
\newcommand{\nnn}{{\mathcal{N}}}

\begin{document}

\title{Coherent states for quantum gravity: towards collective variables}
\author{Daniele Oriti${}^{a}$, Roberto Pereira${}^{a,b}$ and Lorenzo Sindoni${}^{a}$ \\[.5cm]
\small \em a) Max Planck Institute for Gravitational Physics, Albert Einstein Institute, \\
\small \em Am Muehlenberg 1, 14467 Golm, Germany, EU\\
\small \em b) Instituto de Cosmologia Relatividade e Astrofisica, ICRA - CBPF, \\
\small \em  Rua Dr. Xavier Sigaud, 150, CEP 22290-180, Rio de Janeiro, Brazil
}

\date{\small\today}

\maketitle\vspace{-7mm}

\begin{abstract}
\noindent 
We investigate the construction of coherent states for quantum theories of connections based on graphs embedded in a spatial manifold, as in loop quantum gravity. We discuss the many subtleties of the construction, mainly related to the diffeomorphism invariance of the theory. Aiming at approximating a continuum geometry in terms of discrete, graph-based data, we focus on coherent states for collective observables characterizing both the intrinsic and extrinsic geometry of the hypersurface, and we argue that one needs to revise accordingly the more local definitions of coherent states considered in the literature so far. In order to clarify the concepts introduced, we work through a concrete example that we hope will be useful to applying coherent state techniques to cosmology.
 \end{abstract}

\section{Introduction}

Coherent states constitute a key tool to investigate the physics of quantum systems, in both their kinematical as well as dynamical aspects, with special attention to the semiclassical limit. They have first been constructed to describe coherent light \cite{glauber} and since then have been applied to a large variety of physical systems (see \cite{klauder, gazeau} for a review of different aspects and applications). 


In this paper we want to contribute to the discussion of coherent states in the particular context of (loop) quantum gravity, with the objective of defining states which have definite semiclassical properties, as specified below, with respect to extended geometrical quantities, rather than local ones, in terms of the graphs on which general quantum states of the theory are defined, in the sense that they depend on many of the data associated to individual links of such graphs.

Let us briefly summarize the relevant steps of the construction of the states.
First, we use the parametrization of the gravity phase space (intrinsic metric and extrinsic curvature of the slice) used in Loop Quantum Gravity (see \cite{lqg} for a review), obtained by passing from metric variables to Ashtekar connection variables \cite{ashtekar,loop}, and then from point-wise defined fields to their smearing along extended curves and surfaces embedded in the manifold. In particular, the connection is replaced by holonomies along elementary paths in the slice and the conjugate triad is replaced by its flux across elementary surfaces, also embedded in the slice. The final (classical) phase space thus obtained can be parameterized by the set of all graphs embedded in the manifold and a pair of holonomy and flux variables per edge of these graphs (again see \cite{lqg} for details). In the standard connection representation of LQG, holonomies are then quantized as multiplication operators and fluxes as invariant vector fields, giving rise to a well defined kinematical Hilbert space. An alternative (non-commutative) representation is also possible \cite{flux}, in which fluxes act by $*$-multiplication and holonomies act as (functions of) derivatives. The constraints describing the gauge invariance of the system are well defined on this Hilbert space. While the Gauss and (spatial) diffeomorphism constraints can be solved - leading to a Hilbert space spanned by invariant spin networks supported on diff-invariant classes of graphs \cite{lqg} - the Hamiltonian constraint can be defined (up to certain ambiguities), but can not be solved in full generality, at the present stage of development \cite{hamiltonian}. For this reason, work on the dynamics of the theory is now exploring alternative strategies, based on the spin foam \cite{SF} and group field theory ideas \cite{GFT} 

At the quantum level, due to the Heisenberg's uncertainty relations, it is of course impossible to resolve individual points on phase space and the issue becomes that of identifying suitable semi-classical states in the (kinematical) Hilbert space of the theory that replace classical configurations, and peak on a given point in continuum phase space given by the connection and the dual triad fields on the 3-slice. 


Given the parametrization used in terms of graphs, the notion of peakedness is a bit more subtle than usual, as one is mixing two types of approximations: one of quantum nature, stating that, after quantization, the conjugated variables describing the connection and the triad fields should have minimal (but nonzero) spread; and then one of classical nature due to the discretization of the original continuum phase space needed for the construction of the Hilbert space.

In addition to the above, there is the issue of diffeomorphism invariance and background independence of the theory, a defining feature of gravitational theories. With respect to which scale one is defining quantum coherence, given that scales are dynamical in such a theory? 

This question originates from the constrained nature of the gravitational phase space. Let us discuss this in more detail. As well-known, there are three sets of constraints describing the gauge invariance of the phase space: the Gauss, 3d-diffeos and the Hamiltonian constraints. Classically, the kinematical continuum phase space is described by connections and electric fields. One could reduce already at the classical level w.r.t. the Gauss constraint, and that would lead us back to the original ADM variables. Reduction w.r.t. the diffeo constraints is more difficult, and a solution to this would amount to an exact characterization of superspace \cite{misner, giulini}. 

Still at the classical - but discrete - level, the phase space is labelled by embedded graphs and a pair of connection and flux per edge of these graphs. A comparison of degrees of freedom becomes non trivial, since the support of the fields has changed from points to graphs. 
At this discrete level, one could as well reduce w.r.t. the Gauss constraint, but one usually reduces after quantization. 
The issue with diffeomorphisms is more thorny. Modding out diffeos leads to a description of the phase space labelled by un-embedded graphs, but whose associated classical variables do not have an obvious geometric interpretation in terms of the continuum ones. The quantization does not improve the situation on this point, even though it can be performed and leads then to a Hilbert space given by the projective limit of a direct sum of Hilbert spaces, one per (embedded) graph, and each one given by square integrable functions over copies of $\SU(2)$, one per edge of the graph. In this representation, fluxes act as invariant vector fields. One can then reduce w.r.t. to the Gauss constraint, leading to gauge invariant functions and then w.r.t the 3d-diffeos by group averaging over diffeo transformations.

The question now is at which level one is interested in constructing coherent states. The limitations come mostly from the classical theory. A coherent state peaks on a certain point of the classical phase space, and one needs first to choose which classical phase space to work with: the continuum one or the one associated to the discrete graphs embedded in the continuum manifold. Since we understand better the degrees of freedom in the continuum setting we choose this classical phase space. Then, since the only constraint we can actually solve and characterize geometrically at this level is the Gauss constraint, the only question we pose at this stage is to construct coherent states that are gauge invariant or covariant w.r.t. the Gauss constraint. We are not aware of any construction of coherent states that uses diffeo-invariant quantities only and that defines coherence in a diffeo-invariant way, that is, that works on the diffeo-invariant Hilbert space, due to the above-mentioned difficulties. So, we will also deal uniquely with diffeo-covariant (as opposed to invariant) coherent states.

Because we choose to construct coherent states peaking on points of the continuum phase space, but still defined on the phase space coming from the discrete variables associated to graphs, we need to understand this approximation. The main strategy to understand it was introduced by Ashtekar, Rovelli and Smolin in \cite{weave}, where the authors construct states weaving semi-classically a certain 3-geometry. These weave states are not coherent states, since they are eigenstates of the intrinsic geometry, thus completely spread over the extrinsic variables. However, the general procedure and construction, as well as the characterization of the nature of the approximation introduced by the graph is valid and we import it in our own construction. Using the tools introduced by Hall \cite{hall}, the weave states were later generalized by Sahlmann, Thiemann and Winkler \cite{gcs, stw, ST}, and later works \cite{BahrThiemann} to proper coherent states on the full kinematical phase space of GR, that is, including the extrinsic geometry, while keeping the essence of the argument. The coherent state used in those papers was defined in \cite{ashetal} as a tensor product of Hall states, one for each edge. We will argue here that this sort of tensor product construction is not optimal when considering collective  observables, which in turn are needed to approximate a discrete set of data by a continuum geometry at macroscopic scales.\footnote{We will often use words like collective variables, extended objects, macroscopic variables. These terms have slightly different meanings, which we will not distinguish. The idea in the background is that we are trying to move from the specification of a state in terms of microscopic degrees of freedom to a specification of a state in terms of coarse grained information. Whether this operation can be connected to an operation of averaging, decimation or change of variables from local to extended ones depends on the specific construction at hand \cite{calzettahu}. In the example that we will consider later, we are going to change the variables with which we parametrize the classical phase space at the discretized level. This is just a convenient example of the general issue of passing from fine grained data to coarse grained data that has to be addressed in the elucidation of the macroscopic properties of any candidate theory for quantum gravity.} 

It is important to highlight the main limitation of the approach, which lies in its purely kinematical nature.
 In the study of the physical properties of a given quantum many body system, the structure of the chosen (approximate) wavefunction has to be optimized in order to appropriately capture the relevant physical features of the system. This operation, while kinematical in nature (it amounts
to a specification of a corner of the Hilbert space in which the dynamics will be explored), cannot
be totally independent from specifications of the dynamics of the system and of the particular regime in which the system is probed (\eg the condensed phase, etc.).

In particular, the factorized wavefunction, while certainly relevant as a state for a many body problem, is not a good choice when considering interacting
particles, for which the eigenfunctions should not be expected to be factorized. So, dynamics is crucial to understand how to proceed in the construction of approximate physical states. We could say that this paper is a first step in this direction.

Also, it is often more convenient to pass to a second quantized description of a many body system, with the effective number of particles (and their phase spaces) becoming a quantity to be specified only as an average and not as an exact property. Such Fock representation is not (yet) available for LQG or GFT. Still, passing to a formalism that is less sensitive to
the fine grained structure of a specific graph,   while remaining able to capture macroscopic geometrical quantities, as we do in this paper, represents a necessary step to address concretely the semiclassical limit of LQG, irrespectively of whether wavefunctions are given in terms of coherent states, or other states\footnote{That this must be the case is obvious when considering the possibility that the dynamics, specified by the Hamiltonian constraint, might involve the action of graph changing operators.}.

Finally, we have defined in a previous work \cite{OPS} an alternative to Hall states as coherent states for a single copy of the group, using a flux representation for LQG \cite{flux}. The states defined there have improved peakedness properties as compared to Hall states (see \cite{gcs}), and we will use them when constructing coherent states for collective observables in the present paper.  

\

The plan of the paper, in more detail, is the following. We start in section~\ref{weave} by critically reviewing the weaving argument under the form given in \cite{stw}. We will identify certain assumptions and propose an alternative. One of the main points we would like to make is that the choice of coherent states made in \cite{ashetal} is too local, if one is interested in collective observables, and we would like to consider a more collective construction adapted to them. Before trying to give a general construction we will work through a concrete example in section~\ref{example} that we hope will be useful to applying coherent state techniques to cosmology. We will then consider the general procedure in section~\ref{collstates}, highlighting the ambiguities and choices made in this work, before concluding with perspectives and possible developments.

\section{Approximation scales and general construction} \label{weave}

Consider the desired semiclassical state $\Psi_{(A_0,E_0)}$, peaking on a given point of phase space, labelled by a certain pair of datasets, $A_0,E_0$ associated, respectively, to the extrinsic and intrinsic geometry.  To properly construct it with the tools of LQG, let us move to the discretized phase space, where, for each graph, connections are replaced by holonomies supported on edges of the graph and inverse triads are replaced by fluxes supported on surfaces dual to the same edges\footnote{This means that, strictly speaking, the definition of the fluxes requires more than the information contained in the graph itself; in fact it requires the specification of a surface dual to the edge of the graph (and a reference point on the graph itself) \cite{lqg}. For simplicity, we do not use explicitly this extra information, but it has to be kept in mind.}. The minimal set of variables to be used, then, consists of a graph $\Gamma$ and a pair $(h^0_e,P^0_e)$ per edge of the graph. The subtlety here is that, with the graphs sampling the geometry up to a given resolution, many graphs represent the same continuum information. Besides this difficulty, it is clear that the exact identification of the variables associated to each graph is both gauge (Lorentz) and frame (diffeo) dependent.

\subsection{Classical preliminaries}

Let us try to make this statement more precise. Consider the phase space point labelled by $g_0:=(A_0,E_0)$. Choose then a graph $\Gamma$ and embed it on the manifold given by the 3-slice, using the data $g_0$ for that. This means for instance that the lengths of edges and areas of dual surfaces are computed using the intrinsic metric $q_0$ obtained from the triad, and this information is encoded in the discrete variables $P^0_e$. The extrinsic curvature part of $g_0$ is encoded in the variables $h^0_e$, the parallel transports along the edges $e$ of the graph.

We see that there is a certain number of ambiguities in the set of variables $(\Gamma, h^0_e,P^0_e)$ describing the continuum phase space. For a general configuration $(A_0,E_0)$ one would need infinitely many graphs to describe all its properties using the discretized variables \footnote{For metrics with a high degree of symmetry, as in mini-superspace approximations, it is conceivable that a single graph - if not a single point as in homogeneous contexts - would capture all degrees of freedom. One should be careful though since in this context one has no control over perturbations around the mini-superspace metric. Those perturbations might be generated as soon as the dynamics is allowed to excite the other degrees of freedom of geometry.}. 
For a fixed graph, the different configurations $(h_e^0,P_e^0)$ label different possible embeddings of this graph on the 3-slice, according to the continuum data $g_0$. 

It is also clear that both ambiguities are correlated. In order to access all the information represented by the metric $g_0$ one could take two attitudes. The first would be, say, to consider only configurations such that holonomy and fluxes take the same value everywhere on the graph (as long as compatibility conditions are properly taken in to account). One would then probe the geometry by considering finer or coarser graphs. This is the attitude taken in the dynamical triangulations program, or the more recent causal version it \cite{cdt}. A second attitude, closer to the original Regge calculus approach \cite{regge}, would be to fix once and for all the graph. The original continuum phase space would then be probed by different possible embeddings, or equivalently, by varying the holonomy and flux variables. One can always take a fine enough graph such that the continuum phase space is approximated to a given precision \cite{hartle}. Let us stress that all this is still at the classical level and regards the identification of the gauge invariant (classical) phase space, setting only the stage for the quantization program.

 The LQG formalism requires a combination of the above, as the full specification of the Hilbert space of the theory involves both a sum over the variables associated to each graph as well as considering all possible graphs embedded in the manifold. 

Suppose then that a graph and an embedding of that graph have been chosen. The continuum metric is then approximated by a finite set of variables $(h^0_e,P^0_e)$. This approximation, classical in nature, is controlled by the characteristic edge length of the embedded graph $\epsilon$. This scale is computed in terms of the intrinsic geometry contained in $g_0$, and it has statistical nature, representing a sort of cutoff with which the continuum geometry is sampled. 

The geometry $g_0$ that we want to approximate will be characterized by many length scales, in general, associated to intrinsic and extrinsic curvatures. Also, the states we are considering here are still kinematical. However, foreseeing a possible use for studying the effective dynamics, we might want to introduce a restriction on the kind of data that we want to describe to the so called {\it nice slices} \cite{niceslices}. Nice slices are the leaves of a particular choice of foliation of a four dimensional geometry in such a way that the intrinsic and extrinsic curvature are everywhere sub-Planckian. In the parametrization $g_0$ for the classical phase space, this would represent the statement that the curvature radius (intrinsic and extrinsic) is much larger than the Planck length. Of course, the possibility of giving a nice slicing ultimately depends on the four dimensional geometry, which have to be low curvature (essentially because of Gauss--Codazzi relations).
Barring special cases of high symmetry, the radii of curvature (intrinsic and extrinsic) will be thought to be much larger than the statistical scale $\epsilon$.

As said, for generic geometries, it is difficult to specify the curvature scale in a unique way. In symmetric configurations this is more straightforward. For example, it would naturally be given by the Schwarzschild radius, for spherically symmetric and stationary black holes. It will suffice to keep in mind that it will represent the scale of the features of the kind of geometries that we want to use for semiclassical states, and that it is a IR scale.

So, we can identify in general three sorts of scales: the first associated with the intrinsic data, measuring the size of the system; the second associated to the extrinsic data measuring the scale in which the system changes in time; and the third associated to an average (4d) curvature scale. When considering a specific example in the following the meaning of these scales will become transparent.

In the end, it is clear that, sampling the given geometrical data with a discrete structure, we introduce a (classical) error which can be estimated to be controlled by the ratio $\epsilon/L$, where $L$ is the smallest of the scales discussed above.


\subsection{Quantum/Semiclassical analysis}

In order to deal with the semiclassical limit, we are interested in constructing a coherent state $\Psi_{(h^0_e,P^0_e)}$. Following the loop quantization, this state lives in the space $L^2(\SU(2)^{\times N})$ of square integrable functions of many copies of the group, one for each of the $N$ edges of graph. One will later consider the gauge invariant space $L^2(\SU(2)^{\times N}/\SU(2)^{\times V})$ defined by taking the quotient w.r.t. the action of the internal group on the $V$ nodes of the graph.  Holonomy observables are quantized as multiplication operators and flux observables as right invariant vector fields on the group. The non trivial commutators between fundamental operators are given by \cite{lqg}:
\be 
\left[ \, \hat{E}_e^i,\hat{h}_e \right] = i \cp  R^i \triangleright \hat{h}  \;\;  ,\;\; \left[ \, \hat{E}_e^i,\hat{E}_{e'}^j \, \right] = i \cp \epsilon^{ij}_k \delta_{e, e'}\hat{E}_e^{k},
\label{hfa}
\ee
where $\cp$ 
has the dimension of a length squared\footnote{We are using units in which $c=1$.}. The fluxes have the dimension of a length squared. This is because the fluxes are constructed out of triad fields, which are dimensionless, just as metrics are dimensionless in our choice of conventions. When integrating over a surface to define the flux, it acquires the dimension of length squared. The connection has a dimension of an inverse length, and can be integrated over an edge giving a dimensionless quantity. The holonomies - which are necessarily dimensionless - are thus naturally defined without the introduction of any new scale. The Immirzi parameter $\gamma$ appears in the commutator because of the choice of configuration variables:

\be
A_a^i := \Gamma_a^i + \gamma K^i_a.
\ee


At the classical level it does not affect the solutions to the equations of motion, being related to a topological term, but it plays an important role in the quantum theory.

Interestingly, in this algebra, the role of $\hbar$, which distinguishes the classical from the quantum regime in standard quantum systems, is thus played by the composite quantity

\be
t:= 8 \pi G_N \hbar \gamma  = \lp^2 \gamma,
\ee
where, in these units,
\begin{equation}
\lp = (8 \pi G_{N} \hbar)^{1/2}
\end{equation}
is the Planck length.

Consequently there are three (fundamental) constants which might be relevant for the definition of the semiclassical limit: $G$, $\hbar$ and $\gamma$, all contributing to the semi-classicality parameter $t$. Coupling with matter \cite{matter} as well as, of course, the quantum dynamics, might disentangle them and distinguish their  respective roles

One could also take dimensionless phase space variables and define:
 \be \tilde{E}_e := E_e / l_p^2, \ee 
which means that the intrinsic curvature is measured in $l_p$ units, which makes the semiclassical parameter dimensionless and equal to the Immirzi parameter: 
\be \tilde{t} :=  \gamma. \ee 
Again, one could redefine the connection dividing by gamma and $\tilde{t}$ would be simply equal to one. This would imply that quantum fluctuations should be of order one for coherent states.

Together with the length scales characterizing the classical continuum phase space point and $\epsilon$ measuring the classical error in replacing a continuum metric by a discrete one, the algebra of operators introduces a quantum scale, measured by $l_p$, in the sense that fluctuations should be of order one in Planck units for semiclassical states. In the end we will be interested in imposing some relative conditions on all these scales.

\subsection{Tensor product states}

Let us take a look at the semiclassical properties of the states defined in \cite{ashetal, stw}. They are defined as the tensor product

\be
\Psi^{t,\epsilon}_{(h^0_e,P^0_e)}=\bigotimes_e\, \psi^{t,\epsilon}_{(h^0_e,P^0_e)}
\ee
of states defined edge per edge, and on each edge $\psi^{t,\epsilon}_{(h^0_e,P^0_e)}$ is the Hall state \cite{hall} peaked on the point $(h^0_e,P^0_e)$, for an edge of length $\epsilon$ and semiclassical parameter $t$.

This is very close to the original weave construction \cite{weave}. It is the simplest choice, given the kind of fundamental variables of the theory, and it is the one adopted by {\it all} coherent states constructed to date, in both canonical and covariant (spin foam) contexts. While the state $\Psi$ depends on all copies of the group and on all labels, the state $\psi$ depends only on the variables supported on a single edge. 

This assumption has important consequences. In order to highlight them, let us consider observables which admit a similar decomposition into observables associated to the edges of the same graph; for example, one can consider extensive observables supported on a given (embedded) surface on the 3-slice, which will, in general, intersect many edges of the graph. That is:
\be\label{extensive}
\hO(S) = \sum_e\, \hO_e,
\ee
where the sum is over all the edges $e$ intersecting the surface $S$ (and assuming a single intersection per edge), and where each $\hO_e$ is constructed out of the fundamental operators $\hh_e$ and $\hP_e$ only. For semi-classical states the expectation value is supposed to match the classical (discrete) value $O_{cl}$, and considering expectation values with respect to the above type of states:
\be
O_{cl}\,=\,\mean{\hO} = \sum_e\, \mean{\hO_e}.
\ee
 However,  the chosen forms for the observables and state imply that the squared fluctuations $\Delta O := \langle \hO^2 -\langle \hO \rangle^2\rangle$ are given by the sum of the fluctuations for each edge observable:
\be
\Delta O(S) = \sum_I\, \Delta O_I.
\ee

We see that the quantum fluctuations grow with the size of the system, i.e. with the number of edges involved in the construction of the state and of the observable, and hence the state can not be coherent w.r.t. the extensive operator $\hat{O}(S)$ (and its canonically conjugate variable) unless the semiclassical parameter $t$ is taken to be arbitrarily small, which is however, not allowed by our definitions, being fixed to be of order one. 


\subsection{Towards collective variables}

We would like to consider a different construction based on the notion of collective variables. The strategy behind the construction is that semiclassical states should be written in a way that is adapted to the observables one wants to approximate semiclassically. If one is interested in global properties of a dynamical system with a large number of degrees of freedom, then the states describing those properties semiclassically should be optimized for those variables\footnote{There is no generic prescription for a semiclassical state that works for every observable. As in the case of standard coherent states which are associated to creation and annihilation operators, the optimization holds only for observables that are polynomial functions of the same creation and annihilation operators.}.

\

Let us at this point fix the collective observable we are interested in. Consider then the total electric flux $E(S)$ computed on the surface $S$. In terms of the edge operators it is written as in \Ref{extensive}:
\be
\hat{E}(S)=\sum_I\, \hat{E}_I,
\ee
where $I$ again runs over the intersections between the surface $S$ and the graph. This is a very natural observable based on the flux-holonomy representation. To simplify we consider the dual surfaces $S_I$ to which the $\hat{E}_I$ refer, that is the surface $S$ is the union of the elementary surfaces $S_I$. This avoids the staircase problem, as explained in the appendix of \cite{stw}. We will come back to this later.

Another subtle, and important, point is that the fluxes $E_I$ we want to add need to be defined with respect to the same reference point. The classical observables corresponding to $E_e$, such that the flux-holonomy algebra holds, are given by the following expression \cite{qsd7}:
\be\label{flux}
E_e^i := tr\left( \frac{ \sigma^i}{2} h_e(0,p_e)\left[ \int_{S_e} h_{\rho_e (x)} * E(x) h^{-1}_{\rho_e (x)}\right] h^{-1}_e(0,p_e)\right).
\ee
The elements in this definition are the following. $\rho_e(x)$ denotes a path in $S_e$ from point $p_e:=S_e\cap e$ to the point $x_e \in S_e$, and belonging to a system of paths to be chosen from the start. The definition is such that the flux is covariant under internal gauge transformations: $E_e \rightarrow g(0) E_e g(0)^{-1}$, where $g(0)$ is the group acting on the source of the edge (the reference point). 

In defining the total flux we should be careful that it is also gauge covariant and should transform the expression above such that all elementary fluxes transform on the same reference point. The simplest way is to consider the quantity:
\be\label{flux2}
E_{e , x_0}^i := tr\left( \frac{ \sigma^i}{2}\left[ \int_{S_e} h(x_0\rightarrow x) * E(x) h^{-1}(x_0\rightarrow x)\right] \right),
\ee
where $h(x_0\rightarrow x)$ is the parallel transport around a path from an arbitrary reference point $x_0$ to $x\in S_e$. In this way any flux will transform with the same group element, being gauge variant quantities specified in the same frame. Therefore it is also possible to define the total flux in a consistent way\footnote{\label{footflux}Note that one could take a similar definition:
\be
E_{e , x_0}^i := tr\left( \frac{ \sigma^i}{2} h(x_0\rightarrow p_e)\left[ \int_{S_e} h_{\rho_e (x)} * E(x) h^{-1}_{\rho_e (x)}\right] h^{-1}(x_0\rightarrow p_e)\right),
\ee
which is close to the previous definition for very fine graphs. Both give similar results and the difference in the definition could be reabsorbed in a redefinition of the system of paths. The second definition is more directly related to the elementary fluxes.}. It should be clear that the definition of the fluxes depends on a number of structures: the system of elementary dual surfaces $S_e$, the system of paths and the reference point. A natural system of paths is to take always geodesics between two points, as defined by the intrinsic metric one wants to approximate. Also, dual surfaces are naturally defined. The choice of reference point is more ambiguous, as we shall see later. We will come back to this point later. For the purposes of the present discussion, we will assume that a reference point and a system of paths have been defined globally, in such a way that all the ambiguities due to the gauge symmetry
of the system are reduced to the minimum. This is not completely equivalent to a choice of gauge, given that, even with a fixed reference point, a different choices of paths does not lead to a
gauge transformed flux, for instance, given that the connection, in principle, is not a pure gauge.
The dependence on the system of paths is an intrinsic and unavoidable ambiguity of this formalism.

\

The next step is to define the conjugate variable to $E(S)$. In principle and up to certain ambiguities related to the fact that we work with a discrete phase space, the conjugate variable should be determined from the algebra of operators. Since the algebra \Ref{hfa} is not in canonical form, we need to do some preliminary work before defining the conjugate variable. 

Some results in this direction were reported in \cite{OPS} where we defined variables on the phase space corresponding to a single edge that are close to canonical. The idea was to work in the flux representation \cite{flux} as this is more natural for the kind of observables we are interested in. We were able to construct a state with the following semiclassical properties. First, the expectation value of the flux operator is exactly equal to the classical value, that is, for a normalized state $\psi^t_{h_0,y_0}\in L^2(SU(2))$ peaked on the phase space point $(h_0,x_0)\in \SU(2)\times \su(2)\sim SL(2,\C)$, we have:
\be\label{expf}
\langle \psi^t_{h_0,y_0} | \hat{E}^i | \psi^t_{h_0,y_0} \rangle = y_0^i.
\ee
Note that this is only true for Hall states \cite{gcs} at the zero-th order in $t$. A similar result for holonomy seems impossible, due to the non-abelian nature of the group, even though we do not have a proof of this. The best we can do is to search for coordinates on the group manifold $\varphi^i(h)$ such that the expectation value is as close as possible to the classical value. Together with this we ask that the commutation relations $[ \hat{E}^i, \varphi^j(\hat{h}) ]$ are close to $\delta^{ij}$. One can see that both issues are tied together \cite{OPS}. The main results are the following:
\be\label{exph}
\langle \varphi^i(\hat{h}) \rangle_{\psi_0} = \varphi^i(h_0) + O(P^2_{h_0})
\ee
and
\be\label{commut}
[ \hat{E}^i , \varphi^j(\hat{h}) ] = it \left( \delta^{ij} + O(|P_h|^2)\right).
\ee
In the expressions above, $P_h$ is defined as 
\be
P^i_h:= -\frac{i}{2} \text{tr} (h\sigma^i)
\ee
and $\varphi^i(h)=f(P_h^2)P^i_h$, with
\be\label{fx}
f(x)=1+\frac{3}{10}x + O(x^2).
\ee
The above expressions imply:
\be
\langle [ \hat{E}^i , \varphi^j(\hat{h}) ] \rangle_{\psi_0} = it \left( \delta^{ij} + \epsilon^{ij}{}_l f(P^2_{h_0} ) P^l_{h_0} + O(P^2_{h_0})\right).
\ee
All approximations are controlled by the distance of the the group element $h_0$ to the identity. Notice that the logic here is very different from the one in \cite{gcs,stw}, where semiclassical approximations were controlled by $t$. We think of $t$ as of order one and can not be used to control any approximation. Note also that if the group were to be abelian - the abelian limit $\U(1)^{\times 3}$ of $\SU(2)$ is often used in the literature - the expressions above would be exact. We see that the non abelian case is qualitatively different. This will become clear in the next section through a concrete example.

Now that we have the conjugate variable to an elementary flux, we can proceed to define the conjugate to the total flux $E(S)$. For simplicity, one would start with something of the form:
\be
\Phi^j(S) := \sum_I \, c_I \, \varphi^j(h_I),
\ee
where the sum is again over intersections and $c_I$ are arbitrary complex coefficients. Using the results discussed above, we have that
\be
[E^i(S),\Phi^j(S) ] \approx \left(\sum_I c_I\right) \delta^{ij}.
\ee
We need then that $\sum_I c_I = 1$, so that typically, $c_I  \sim 1/N$, where $N$ is the total number of intersections\footnote{There is an intrinsic ambiguity in this definition, at least at this level of the analysis, resulting in the impossibility of giving a specific prescription for the coefficients $c_I$. Indeed, without a full specification of a canonical transformation from the phase space parametrized in terms of holonomies and fluxes to a parametrization in terms of other (discretized) canonical coordinates, of which the total flux is one of the variables, it is impossible to resolve this ambiguity.}.

Notice that this definition of $\Phi$ implies that it is an {\it intensive observable}, as it should be expected from the fact that it is the conjugate variable to an extensive variable (as it is the electric flux).

Even though this observable is naturally defined from the algebra of fundamental operators, its classical interpretation is not as clear, as it is not a familiar variable in lattice gauge theory (see for instance \cite{wilson}). Moreover the expression we obtained holds only at the linear approximation. To go for the appropriate approximation in general field configurations, we need to extend the notion of coordinate canonically conjugate to the fluxes, and disentangle the nonlinearities. For instance, if  we consider the situation with only two links being relevant, the following three functions:
\begin{equation}
\frac{\varphi^{i}(h_1) +\varphi^i(h_2)}{2}, \qquad \varphi^i\left((h_1h_2)^{1/2}\right) ,
\qquad \varphi^i\left((h_2h_1)^{1/2}\right)
\end{equation}
all have the same linearized approximation, for group elements close enough to the identity, and are thus equivalent to the quantity we defined above.

As for the fluxes we need to make sure that $\Phi^j$ is covariant under gauge transformations.  It is sufficient that each $h_I$ transforms in the same way, and with respect to the same reference point $x_0$ used to define the fluxes $E_I$, such that the algebra is preserved as well. We take then for each edge, the Wilson loop \cite{wilson} starting at $x_0$ containing the edge and coming back to $x_0$. Denote it $h_{I, x_0}$. It transforms as $h_{I, x_0}\rightarrow g_{x_0} h_{I, x_0} g^{-1}_{x_0}$, which implies that $\varphi^i(h_{I, x_0})$ transforms as a vector.

\

Having dealt with the observables we can now construct the state. We will use the state defined in \cite{OPS} on a single copy of the group as a template for the collective state. Its precise definition, given in the flux representation, is:
\be
\psi^t_{h_0,y_0}(y) := K^t(y-y_0)\star e^{\frac{i}{t}P_{h_0}\cdot y}.
\ee
We refer to \cite{flux,OPS} for the definition of the star product used in this expression. In the holonomy representation the same state is written as\footnote{Not to overload notations, we consider a state on $\SO(3)$ only. For the extension to $\SU(2)$, see again \cite{OPS}.} 
\be
\psi^t_{h_0,y_0}(h)=K^t_{h_0}(h)e^{-\frac{i}{t}P_{h}\cdot y_0}.
\ee
In the expression above $K^t_{h_0}(h)$ defines a gaussian on the group peaked on $h_0$ and can be taken to be the heat kernel on the group. We would like to emphasize that the particular choice of coherent state is not so important for the point we want to make in this paper, even though the state just defined is particularly well suited for observables based on fluxes. Other global observables, such as the area of a surface or the volume of a region will probably require different states. However, any other construction should satisfy properties similar to (\ref{expf}-\ref{commut}).

Coming back to the collective state, it would be simpler to have a pair of collective variables living in $\SU(2) \times \su(2)$. The total flux is already in the algebra. To get an holonomy one needs to invert the map $\varphi: \SU(2) \rightarrow \R^3$, and this is always possible in a certain neighborhood of the identity. So denote $h_0(S):=\varphi^{-1} (\Phi(S))$. Noticing also the expansion \Ref{fx} of $\varphi(h) = P_h + O(|P_h|^3)$, we have that $P_{h_0(S)} \approx \Phi(S)$. We then take our state to be defined in the flux representation as:
\be
\Psi_{E(S),\Phi(S)}(\{ x_e \}) := K^t \left(\sum_I x_I - E(S)\right) \star e^{\frac{i}{t}\Phi(S) \cdot (\sum_I x_I)} \otimes \psi_{rest}(\{ x_e \}),
\ee
where we take $K^t$ to be the heat kernel in Fourier space. $\psi_{rest}$ describes the degrees of freedom other than $E(S)$ and $\Phi(S)$, and that are necessary to fully specify the geometry of the hypersurface. Notice in fact that this state is still defined on the Hilbert space associated to the full graph.

By construction this state minimizes uncertainties of the collective operators $\hat{E}(S), \hat{\Phi}(S)$ while it might have larger fluctuations for more local operators, depending on $\psi_{rest}$.

Another important point to notice concerns gauge transformations. As we have stressed, the elementary fluxes, and thus the total flux, have to be computed in the same reference frame corresponding to an arbitrary vertex on the graph. The fluxes appearing in the above state, therefore, correspond to the fluxes obtained after parallel transporting (by means of appropriate combinations of holonomies $h_e$) those initially associated to the various  edges of the graph to the same point. The gauge transformations still act on each vertex of the graph itself. Thus our flux variables, and our state, will transform differently than states defined in terms of the original flux variables (which are expressed each in the frame of one of the two graph vertices touched by their associated edge). In particular, they will not satisfy the standard closure condition at each vertex of the graph \cite{flux}, but an appropriately parallel transported version of it. This difference in appearance of the action of gauge transformations can also be verified, of course, in the group representation, that is when expressing the same state in terms of the group variables conjugate to the parallel transported fluxes, which are in fact not the standard holonomies associated to each edge of the graph, but again a combination of them with the parallel transports needed to go from the initial vertex to the one chosen as common reference point. 

\subsection{An analogy: a system of many point particles in flat space}
To clarify the discussion of the LQG case, we might draw an analogy with a much simpler system. In the case of several point particles moving in a flat Euclidean three dimensional space,
an obvious parametrization of the phase space is in terms of the coordinates
and the momenta of the various particles. 
If convenient, one can equivalently parametrize the same phase space with a different choice of variables, by means of a canonical transformation.
The significance and the specific form of the canonical transformation might reflect symmetry properties, or specific features of the dynamics.

If the classical system is characterized by an Hamiltonian where the interaction depends only on the relative positions, it makes sense to choose canonical coordinates which
distinguish the center of mass, which describe the overall motion of the system (highlighting the global conserved quantities like total momentum, and total angular momentum), from the relative coordinates, that control the fine details of the dynamics of the system.


As canonically conjugate collective observables, thus, we can consider the pair (Q,P) of the center of mass coordinate and total momentum. In a certain sense it is a coarse grained information. To preserve the canonical commutation relations, because
the total momentum is an extensive variable, \ie it is scaling with the number of particles, the position of the center of mass is an intensive variable, representing an average property.
The choice of the completion of this pair into a full set of canonical coordinates for the total phase space is largely arbitrary, of course, and it has to be motivated on dynamical grounds. This is particularly delicate when considering the quantum theory. Indeed, when considering a 
state which is constructed as a tensor product of single particle wavefunctions (barring symmetrizations) certain operators might have optimal properties (minimal fluctuations), while others might be affected by large uncertainties. This is exactly the case we are facing here.

The very nature of the typical dynamics of many particles, involving more or less complicated interactions,  leads to the necessity of using nonfactorized wavefunctions (or even density matrices). Without a proper assessment of the properties of the Hamiltonian constraint, it is impossible to propose one definite class of states as good toy states that will give reasonable approximation to a physical state. However, for the semiclassical regime of quantum gravity it
is reasonable to expect that the relevant states (or density matrices) will have the property of
having good expectation values with small fluctuations for the geometric operators associated to
more global geometrical properties, with possibly large fluctuation for small scale details of the geometry (space-time foam). This analogy explain the motivation and goal of our work in the quantum gravity context.

\

In the following we will work through a concrete example that will enable us to get some intuition on the construction presented so far.

\section{Example}\label{example}

It is instructive to consider an explicit case of our construction, in which all the calculations can be performed. This will allow us to further appreciate the advantages of using a given collective state instead of a microscopic state in addressing the behavior of large structures.
We want to construct a semiclassical state for a given geometry in such a way that information about geometry is encoded into extended objects and then compare the situation with the state
constructed edgewise. 

\subsection{Choice of the classical point}
We choose, as a classical phase space point $g_0$, the three manifold whose topology is that of a torus $T^3=(S^1)^3$ with flat three dimensional metric. 

In cartesian coordinates
\begin{equation}
ds_3^2 = q_{ab}dx^a dx^b, \qquad
q_{ab} = a^2(\tau)\delta_{ab}
\end{equation}
with $x,y,z\in [0,L_0/a)$, and $a(\tau)$ the scale factor that might depend on a time parameter $\tau$. The radii of the torus are assumed to be equal to $L_0/a$, such that their physical length is equal to $L_0$. They might be different, in principle. The choice of such a geometry  (see \cite{magliaromarcianoperini} for a different choice) has the advantage of allowing more immediate calculations and comparisons, highlighting the relevant features with the minimal amount of irrelevant technical complications.

With a flat metric we can choose a gauge in which:
\begin{equation}
e^i_{a} =a(\tau) \delta^i_a
\end{equation}

The extrinsic curvature is chosen to be proportional to a Kronecker delta as well.
\begin{equation}
K_{ab} = \dot{a}(\tau) a(\tau) \delta_{ab} =: H(\tau) q_{ab}
\end{equation}
where $\dot{a}(\tau)$ and $H(\tau):=\dot{a}/a$ to suggest a relation to a cosmological scenario in which the whole dynamics is controlled by the scale factor. 

\subsection{Classical observables}

We can then induce the Barbero connection, in this gauge,
\begin{equation}
A^{i}_a = \gamma K_a^i= e^{i b} K_{ab} = \gamma \dot{a} \delta^i_a = \gamma H e_a^i
\end{equation}
and easily compute the parallel transport on a geodesic of the metric. In flat space these geodesics are just straight segments. For the segment connecting two points $A,B$:
\begin{equation}
\vec{x}_A - \vec{x}_B = \frac{\ell}{a} \vec{\mu}_{AB}
\end{equation}
The parallel transport gives:
\begin{equation}
h_{AB} = \po \exp \left(  i  \int A^i_a \frac{\sigma_i}{2} dx^a \right) = \cos\Theta_\ell \id + i \sin\Theta_\ell \mu^j\sigma_j
\end{equation}
\begin{equation}
\Theta_{\ell} = \frac{ \gamma H \ell}{2}.
\end{equation}

For a generic choice of curves and generic connection configurations it is difficult to give the parallel transport in an analytic form as here. However, for the case in which the matrix
\begin{equation}
M =A^{i}_{a}(x) \frac{i\sigma_i}{2} \dot{x}^a,
\end{equation}
is a constant, \ie for the choice of curves and connections such that
\begin{equation}
A^{i}_a(x) \ddot{x}^a + \partial_b A^{i}_a \dot{x}^b\dot{x}^a=0,
\end{equation}
an analogous expression holds. This equation, in some sense, can be seen as an equation for autoparallel curves for the connection. In this particular case, the geodesics are also autoparallel for this connection.

The connection is associated to a second length scale, $L_H = H^{-1}\gamma^{-1}$, with nice slicing implemented automatically, giving that $L_H \gg \lp$.

Notice that, to simplify even further the analysis, we could set $H(t)=0$ and consider then a chunk of flat spacetime. However, in keeping the explicit dependence on the connection we will also have the chance of appreciating a number of subtle issues that are often overlooked.

The connection has a nontrivial field strength
\begin{equation}
\mathbf{F}_{ab} = \frac{\sigma_i}{2} \left(\partial_a A^{i}_b - \partial_b A^i_a + \epsilon_{ijk}A^{j}_a A^{k}_b\right)
 = M \sigma^i \epsilon_{iab},
\end{equation}
where
\begin{equation}
M= \frac{1}{2}\gamma^2 H^2 a^2
\end{equation}
which means that the connection is not a pure gauge, and that the closed holonomies will be in general different from the identity. 

To compute geometrical quantities relevant for the LQG representation, we will have to integrate two-forms over plaquettes. However, these two forms will be Lie-algebra valued, and hence will require delicate manipulations to ensure that the integrated quantities will transform well under gauge transformations.

Consider the form
\begin{equation}
\mathbf{B}=B^{i}_{ab} \sigma_i dx^a\wedge dx^b,
\end{equation}
whose transformation law under gauge transformation is
\begin{equation}
g^{-1}(x)\mathbf{B} g(x)
\end{equation}

Under a gauge transformation, the connection will transform as
\begin{equation}
\mathbf{A}' = g^{-1} \mathbf{A} g + i g^{-1} (\partial g)
\end{equation}
while the parallel transport will become
\begin{equation}
h(x_0 \rightarrow x) \rightarrow h'(x_0 \rightarrow x) = g(x)^{-1} h \, g(x_0)
\end{equation}

Given the surface of integration $S$ and a reference point $x_0$, construct a given system of paths connecting each point of the surface $x\in S$ to the reference point $x_0$, and define the
integrated form as
\begin{equation}
\mathbb{B} = \int_S h(x\rightarrow x_0) \mathbf{B} h(x_0\rightarrow x)
\end{equation}

As a consequence of the transformation properties of the various terms of the integrand, one easily realizes that
\begin{equation}
\mathbb{B} \rightarrow g^{-1}(x_0) \mathbb{B} g(x_0)
\end{equation}
under a gauge transformation.

In the specific case we are interested in, we will consider integrals of the following form
\begin{equation}
\mathbf{B} = f(\tau) \sigma^a \epsilon_{abc} dx^b\wedge dx^c
\end{equation}

Consider the surface $S$ as the plaquette of coordinate size $\epsilon/a(\tau)$, laying on the $z=0$ plane, and centered in the origin of a cartesian reference frame. If we consider a general position of the reference point, and a system of paths connecting it to the surface as made of straight segments, we can  compute
\begin{equation}
\mathbb{B} = f(\tau) \int_{-\epsilon/2a}^{\epsilon/2a} dx\,dy\, \left( \cos(\Theta(x))\id - i \sin(\Theta(x))\mu^a\sigma_a\right)\sigma^3\left( \cos(\Theta(x))\id + i \sin(\Theta(x))\mu^a\sigma_a\right).
\end{equation}
Things are easily computed if we place the point $x_0$ on the origin of the cartesian coordinate system.
Indeed, in this case, $\mu^a = (\cos{\xi},\sin\xi,0)$, and by symmetry
\begin{equation}\label{B}
\mathbb{B} = \frac{ f(\tau)}{a^2} F(\gamma H,\epsilon) \sigma^3,
\end{equation}
where
\begin{equation}
F(\gamma H,\epsilon)  = a^2 \int_{-\epsilon/2a}^{\epsilon/2a} dx\,dy\,\cos(2\Theta(x))
\end{equation}
For small values of $\epsilon$, one sees that
\begin{equation}
F(\gamma H,\epsilon) \approx \epsilon^2 \left( 1 - \frac{1}{3} \epsilon^2 \gamma^2 H^2 \right)
\end{equation}
Of course, moving around the reference point will result into a less straightforward expression. However, for small values of the dimensionless parameter $\epsilon^2\gamma^2 H^2=\epsilon^2/L_H^2$, 
\begin{equation}
F\approx \epsilon^2
\end{equation}
and the form
\begin{equation}
\mathbb{B} \approx \frac{f(\tau)}{a^2} \epsilon^2
\end{equation}
which coincides with the result for an abelian group. The case of the electric flux (our main interest here) corresponds to the choice $f(\tau) = a^2$. Another quantity of interest is the magnetic flux on the surface, that is, the integral of the field strength of the connection. It corresponds to the choice $f(\tau) = a^2 \gamma^2 H^2$.

Let us now move the reference point around. Consider a reference point still on the plane $z=0$, but far away from $S_e$. It is convenient to use the second definition of fluxes given in footnote \Ref{footflux}. We need to compute
\be
h(x_0 \rightarrow 0)\,  \mathbb{B} \, h(0 \rightarrow x_0).
\ee
Inserting \Ref{B}, we get
\eqa
\mathbb{B}_{x_0} &\varpropto&  \left( \cos\Theta(x_0)\id - i \sin\Theta(x_0)\mu^a\sigma_a\right)\sigma^3\left( \cos\Theta(x_0)\id + i \sin\Theta(x_0)\mu^a\sigma_a\right) =\nonumber \\
&=&  \sigma^3 \left( \cos\Theta(x_0)\id + i \sin\Theta(x_0)\mu^a\sigma_a\right)^2 = \sigma^3 \left( \cos 2\Theta(x_0)\id + i \sin 2\Theta(x_0)\mu^a\sigma_a\right). 
\neqa
From the last expression we can read the components of $\mathbb{B}$:
\be 
B_{x_0}^3 \approx  \frac{f(\tau)}{ a^2} \epsilon^2  \left( 1 - \frac{1}{3} \epsilon^2 \gamma^2 H^2 \right) \left( 1-2\Theta^2(x_0)\right) \approx \frac{f(\tau)}{ a^2} \epsilon^2 \left( 1 - \frac{1}{3} \epsilon^2 \gamma^2 H^2 - \frac{1}{2}l^2_{x_0} \gamma^2 H^2 \right)
\ee
and 
\be
B_{x_0}^b \approx \frac{f(\tau)}{ a^2} \epsilon^2  \left( 1 - \frac{1}{3} \epsilon^2 \gamma^2 H^2 \right)\mu^a \epsilon_{3ba} l_{x_0}\gamma H,
\ee
for $a,b=1,2$. We see that approximations are controlled by both by $\epsilon/L_H$ and $l_{x_0}/L_H$.

The total flux on the $z$ direction is then given by:
\be
E^3(S) \approx \sum_I \, \epsilon^2 \left( 1 - \frac{1}{3} \frac{\epsilon^2}{L^2_H} - \frac{1}{2} \frac{l^2(x_0-p_I)}{ L^2_H} \right)
\ee
up to higher order corrections in both $\epsilon/L_H$ and $l(x_0-p_I)/L_H$ and the other components are small compared to it. Taking the reference point to lie on the middle of the surface makes $l(x_0-p_I)$ of the order of the total size of the system $L(S)$ so that the relevant ratio is $L(S)/L_H$, that is, the size of system compared to the typical scale controlling its evolution in time. Taking a fine enough graph, the correction in $\epsilon/L_H$ becomes negligible.

Let us now compute the conjugate variable $\Phi^j(S)$. As explained in the last section it is given by the average coordinate of Wilson loops going from $x_0$ to the edge and then back to $x_0$. 

Consider then a generic reference point with coordinates $(x_0,y_0,z_0)$ and a link parallel to the z axis, parametrized by
\begin{equation}
\vec{X} : [0,1] \rightarrow T^3; \qquad \tau \mapsto \left(
\begin{array}{c}
(2n+1)\\
(2m+1) \\
(2\tau-1)
\end{array}
 \right) \frac{\epsilon}{2a}
\end{equation}

Therefore, the full holonomy has to be computed as
\begin{equation}
h_{e,x_0} = h(e(1)\rightarrow \vec{x}_0)h(e)h(\vec{x}_0\rightarrow e(0))
\end{equation}

Denoting:
\begin{equation}
n^{i} = (0,0,1)
\end{equation}
and
\begin{equation}
v^{i}_{+}(e) , v^{i}_{-}(e) 
\end{equation}
the unit connecting vectors from the reference point to the tip $(+)$ and tail $(-)$ of the link $e$ which has been assumed to be oriented from the negative to the positive values of the coordinate $z$.
Similarly $d_+$ and $d_-$ will represent the physical distances. To simplify the calculation, suppose further that the reference point lies in the $y=0$ plane and is equidistant from the two tips of the edge, so that $d_+=d_-$. Keeping terms of second order\footnote{{Notice that, in the case of noncompact slices, there is no obvious way to motivate such a truncation for a generic reference point: if the reference point lays many Hubble radii from the edge considered, there is no way in which the parallel transports to the reference points are close to the identity. }} in $\Theta_\pm = \Theta$ and $\Theta_\epsilon$, we get: 
\eqa
h_{e,x_0} &\approx&
\left((1-\Theta^2/2)\id - i \Theta v^{i}_{+}\sigma_i\right)
\left((1-\Theta^2_\epsilon/2)\id + i \Theta(\epsilon) n^{i} \sigma_i\right)
\left((1-\Theta^2/2)\id + i \Theta v^{i}_{-}\sigma_i\right) \nonumber \\
&\approx& \id + i  \left( \Theta_\epsilon n^{i} - \Theta v^{i}_{+} + \Theta v^{i}_- \right)\sigma_i +i\sigma_y \Theta \Theta_\epsilon.
\neqa
Since we work in a flat space the linear term is identically zero and we are left with:
\be
h_{e,x_0} \approx \id +i\sigma_y \Theta \Theta_\epsilon = \id +i\sigma_y \frac{l_{x_0} \epsilon}{4L_H^2}.
\ee
This is an interesting result. Remember that the state we chose to work with is a good semiclassical state as long as the Wilson loops are not too far from the identity. We see from the above expression that this approximation is controlled by the ratio $L(S)\epsilon/L_H^2$, where again we take $l_{x_0}$ to be of the characteristic size of the surface. We see that for situations where $L(S)$ is comparable to $L_H$, $\epsilon$ has to be very small, which means one should take a very fine graph. On the other hand, for situations where $L(S)$ is already much smaller than $L_H$, $\epsilon$ can be taken to be of the order of $L(S)$, which justifies taking a very coarse graph. 


 
To get a better intuition on the scales discussed above, let us see how they look like in a cosmological scenario. Consider then a cosmological model describing the evolution of the scale factor $a(\tau)$. The approximation discussed above is based on two ratios, $\epsilon/L_H$, controlling the continuum limit and $L(S)/L_H$. $L(S)$ is equal to the scale factor times a constant coordinate scale $l_0$, that can be taken to be the size of the system if the scale factor is normalized as $a_0=1$. $L_H= (\gamma H(\tau))^{-1}$, such that for $\gamma\sim 1$, $L_H \sim (\dot{a}/a)^{-1}$ and the ratio is of order $\dot{a} l_0$. We see that $\dot{a}(\tau)$ controls the behavior of this ratio in time. 

\begin{figure}[here]
\includegraphics[width=8cm]{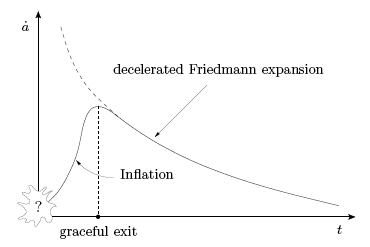}
\caption{Standard inflationary model. Quantum gravity should be relevant at the question mark. (Picture taken from Mukhanov's book {\it Physical Foundations of Cosmology})}
\label{inflation}
\end{figure}

Consider moreover that the universe goes through an inflationary phase \cite{mukhanov}. The behavior of $\dot{a}$ during inflation is quite generic and is such that it starts very small at the beginning of inflation then grows exponentially before re-collapsing to small values for very late times (see figure \ref{inflation}).

Following the discussion above we see that at the beginning of the inflation it is justified to use a coarse graph. The moment when the scale factor starts to grow one is forced to take finer graphs to control the approximations assumed. This conclusion depends of course on a number of hypothesis, first in the construction of the pair of collective conjugate variables, and then on the coherent state, but seems very generic for a Hilbert space based on graphs, as long as the state is based on a single graph. The situation might also change when considering states given by linear combinations of states supported on different graphs.

\subsection{The state}

Having described the classical geometric data that we want to reproduce, let us construct a state along the lines discussed in the previous section. We take a cubic decomposition of the torus, with uniform spacing, say $\epsilon$. The graph is obtained by joining the nodes with straight segments (this is equivalent to the choice of a prescription to color the edges). We could choose different paths, and this would result into the same topological graph, with a different coloring of the edges. 
Notice that this would not be associated to a displacement of the nodes, but rather to a certain choice of the embedding of the edges.

Let us consider an adapted cartesian system of coordinates on the torus, by choosing the origin
in such a way that each edge of the graph is parallel to one of the axis, and in such a way that
the origin sits in the middle of one of the cubes having the graph links as edges.

Take, for instance, the four edges of the cube around the origin, parallel to the $z$ direction. They
will be endowed with certain holonomies, as we said, and with certain fluxes. The fluxes will be computed across square plaquettes on the $xy$ plane, with side $\epsilon$. 

Alternatively, we can store the information about the fluxes into a state that is coherent for the flux
across the union of the four plaquettes that we are considering.

To establish an appropriate comparison, the two states must refer to fluxes computed with respect
to the same reference point and with respect to the same system of paths. For simplicity we choose to work with the system of paths consisting of straight segments connecting the origin to each point of the plaquettes, with the origin being the reference point for all the fluxes. The flux across the surface consisting of the four plaquettes is
\begin{equation}
E^{tot} = F(\gamma H, 2\epsilon) \sigma^3
\end{equation}
as we have already shown. The small plaquettes, instead have fluxes (one has just to chop the integral into the four quadrants, after rescaling) corresponding to
\begin{eqnarray} 
E^{(1)} =  \left(\frac{F(\gamma H, 2\epsilon) }{4} \sigma^3 + (+\sigma^1 + \sigma^2)C \right)\\
E^{(2)} = \left(\frac{F(\gamma H, 2\epsilon) }{4} \sigma^3 +  (-\sigma^1 + \sigma^2)C \right)\\
E^{(3)} = \left(\frac{F(\gamma H, 2\epsilon) }{4} \sigma^3 +  (-\sigma^1 - \sigma^2)C \right)\\
E^{(4)} =  \left(\frac{F(\gamma H, 2\epsilon) }{4} \sigma^3 +  (+\sigma^1 - \sigma^2)C \right)
\end{eqnarray}
where
\begin{equation}
C := \iint_0^{\epsilon} \sin(2\gamma H \sqrt{x^2+y^2}) \frac{x}{\sqrt{x^2+y^2}} \, dx dy
\end{equation}

The four plaquettes are numbered after the quadrants of the plane going counterclockwise.

In this reference system, and with the choice of the same reference point, in particular, it is immediate to realize that:
\begin{equation}
E^{tot} = E^{(1)}+E^{(2)}+E^{(3)}+E^{(4)}
\end{equation}
and this relation can be promoted even at the level of operators, given that the parallel transports are already included in the definition.


Let us now compare our collective states with the standard tensor product states.

At the quantum level, one can now construct two kind of states. One is adapted to the graph, treating the information associated to it edgewise, with a factorized wavefunction, and the other one is adapted to larger structures corresponding to the collective observables we defined. Working in the flux representation, where $x_i$ are Lie algebra variables, associated to the links, these two types of states take the form:
\begin{equation}
\psi^{old}(x_1,x_2,x_3,x_4,\{x_{rest}\}) = \left( \prod_{i=1}^4 \psi^{t}_{h_i} (x_i-E^{(i)}) \right) \times \psi(rest)
\end{equation}
\begin{equation}
\psi^{new}(x_1,x_2,x_3,x_4,\{x_{rest}\}) = \psi_{\Phi^{tot}}^{t}(x_{1}+x_{2}+x_{3}+x_{4}-E^{(tot)} )\times \psi(rest)
\end{equation}
Note that in the second state, (rest) might still contain $x_1,...x_4$, whenever the corresponding dual plaquettes are involved. Once more, the four Lie algebra elements in the new wavefunction can be composed only because they are associated to fluxes defined with respect to the same reference point, and hence they transform in the same way under gauge transformations. In a factorized state, even though it is expressed in the same variables, this is not obvious. The same remark we made in the previous section regarding gauge transformations and their non-standard expression in these variables, due to them being parallel transported to the same point, applies of course here.

It is immediate now to see that the factorized coherent states, while providing the same value for the expectation value of the flux, give a fluctuation around the average that is $4t$ rather
than the value, $t$, that is characterizing the fluctuation of the flux with the adapted coherent state.
Notice that the precise form of the coherent state $\psi$ is not important to reach this conclusion, though
the coherent states introduced in \cite{OPS} are particularly suitable for approximating the chosen observables.

If instead we measure the fluctuation of the individual fluxes from the second state, we will have
the contribution of the state associated to the surface we are considering, as well as the one
of all the other components of the state, $\psi(rest)$, containing a dependence on the link $i$.
Hence, we should expect microscopic geometric information to be reproduced on average, but with
large fluctuations.

The procedure can obviously be generalized for different, larger surfaces and sets of plaquettes, without introducing new concepts (albeit the calculations might be complicated). In particular, the
result about the fluctuations remains the same.

\section{Collective states}\label{collstates}

The previous explicit example has highlighted a number of crucial steps and features that have to be taken care of in the construction of a general semiclassical state associated to a given graph and, in particular, of nonfactorized states designed to support information of the geometry of extended regions of space, made out of many fundamental degrees of freedom. We recapitulate them here, outlining the general construction.

\subsection{Construction of the state}

In the example that we have considered, we have gone through the following steps.

(i) Replace the continuous metric by a discrete one, by first choosing a graph and an embedding. This limits the possibility of resolving the variables $(h^0_e,P^0_e)$  up to a certain precision $\epsilon /L$, where $\epsilon$ is the characteristic discretization scale of the graph (in the embedding metric $h_0$) and $L$ is the characteristic curvature scale of the same metric;

(ii) Choose the surface $S$, a reference point and a system of paths with which all the gauge variant quantities have to be computed. 
Compute the classical values for the electric flux $E^{i}_{0}(S)$ and for the ``conjugate variable" $\Phi_0^{i}(S)$.

(iii) Finally choose a state, such that the expectation values of the observables $E(S), \Phi$ are in agreement to the classical discrete values, computed with  $(h_0,P_0)$, also up to $\epsilon /L$, and also such that the Heisenberg uncertainty relations are saturated, up to quantum corrections of order $t^2$.

Of course, this algorithm can and should be generalized to different kind of operators, like areas, volumes, and other geometrical properties of more or less extended structures. The choice of electric fluxes considered here is due to the fact that we have nice semiclassical states to deal with these variables \cite{OPS}.


Other geometrical features characterizing the slice can also be associated to the macroscopic scale $L$. It represents a typical curvature scale but it might be also associated to topological features, like the radii of a torus, etc.. 

\

The previous steps can be generalized as follows

i) From the continuum structure associated to a compact slice we can construct a discrete sampling by means of a Poissonian sprinkling \cite{Bombelli}, determined by the volume element of the embedding metric, as it is customarily done for  the construction of causal sets \cite{causalsets}.
The random sprinkling consists of the random selection of points on the manifold, which we assume to be compact for convenience, according to the volume element specified by the intrinsic metric tensor. The density of the sprinkling is determined by the total number of points used. If we assume to sample the geometry with $\nnn$ points, this will amount to partition the manifold (according to the Voronoi procedure \cite{Bombelli}) into $\nnn$ random cells of typical size $Vol/\nnn$. If, for convenience, we assume that the volume we are interested in is of order $L^3$, this amounts to assign to each point a chunk of space of typical volume $L^3/\nnn$. 

With a random Voronoi complex we can construct its associated random graph, which will be the object that we will assume to be the support of our semiclassical state. In turn, $g_0$ will induce certain geometrical data on this graph. 


Concerning scales, the random sprinkling is providing, through the density of points, the definition of the length scale $\epsilon$
\begin{equation}
\epsilon = \frac{L}{\nnn^{1/3}}.
\end{equation}

This scale, $\epsilon$, will be the typical scale of all the geometric operators, but, most importantly, it fixes the accuracy with which we can approximate, at the classical level, the continuum geometry with the discrete data selected by the graph.

In the case of the holonomies, the construction ensures also that any holonomy that is induced on the graph, when considered in the following parametrization
\begin{equation}
h=\cos{\theta} \mathbb{I} + i \sin{\theta} n^{i}\sigma_i ,
\end{equation}
is characterized by
\begin{equation}
\theta \sim \frac{\epsilon}{L} = \nnn^{-1/3}.
\end{equation}

Let us stress again that the nature of the scale $\epsilon$ is purely classical and statistical, and it has nothing to do with the Planck scale or with any other dynamical scale, at least at this stage.
Indeed, being the scale of the sprinkling, it sets an obvious UV cutoff that makes impossible to resolve geometrical structures differing on length scales smallers that $\epsilon$ 
itself. However, $\epsilon$ alone is not describing how we are effectively probing spacetime. Indeed, we are codifying this latter scale in the way in which we choose the surfaces and the other extended objects to construct the semiclassical states. It will be the typical length scale of these extended structure to be really defining the physical coarse graining scale.

The necessity of this kind of random sampling is due to the absence, for generic three dimensional geometries, of symmetries that might lead to obvious definition of graphs that are optimally encoding the required geometric information.

ii) The next step involves the identification of the variables that we want to use for the construction of the state. We can either use the graph edgewise, or rather use collective variables, as in the
example discussed previously.

There, we considered a surface $S$ and the edges of the graphs intersecting the surface. Furthermore, we have to compute the classical value of
the conjugate variable $\Phi^i$, that will encode the extrinsic geometry.

At this stage it is important to mention one problem of the construction of states based on extended
observables. One known problem is the so-called staircase problem \cite{stw}. Given a graph with certain associated geometric data, it is possible to
construct operators $O,O'$ such that, while the states gives the correct semiclassical value, and/small fluctuations
for the operator $O$, it might have an expectation value far off the classical one for $O'$, and large fluctuations.
This depends on whether the operator corresponds to objects that are adapted to the graph or not.

The integrated fluxes have to be defined with respect to a definite choice of paths and reference point, chosen once and for all, so that we can consistently compare the fluxes computed for each subplaquette with the flux attached to the surface $S$. 


iii) Finally, according to the given choice of variables, one can construct a coherent state that
is adapted to this choice. The wavefunction {\it will not be factorized} with respect to the single edges.
A similar procedure should be followed for the other variables, to be included in $\psi_{rest}$.
Besides the numerical value of the classical discrete geometrical quantities, 
one will have to complete the discussion by specifying the maximum number of independent surfaces and fluxes (or other necessary variables) needed to completely specify a semiclassical state for the given assignment of geometric data $g_0$.

The state so constructed will have the property of encoding the classical phase space point with minimal quantum fluctuations around the expectation values. 

The improvement with respect to the situation discussed in section 2 is manifest. 


Of course, this very same procedure is guaranteeing us that, when we construct states in terms only of large structures, the expectation values of the fine grained geometrical operators (\eg the parallel transport on a single link) will instead generically suffer from large fluctuations, as to be expected from other arguments. This type of quantum state (nicely semiclassical for macroscopic geometric quantities, manifestly quantum for microscopic ones) could also be interpreted as a specific encoding of the idea of a spacetime foam.

\section{Conclusions and outlook}
Let us summarize what we have achieved in this paper. 

First of all we have reviewed critically the construction of coherent states in a quantum gravity context, at the kinematical level. We focused on the well-developed  kinematical framework of loop quantum gravity, as a solid example of a canonical quantization of geometry in the continuum. Because this results in quantum states of geometry associated to graphs embedded in the spatial manifold, the issue of approximating continuum geometries with discrete structures comes to the forefront, in addition to the issue of approximation of classical observables by means of quantum states. Similar quantum states, and thus the above issues, appear in several approaches to quantum gravity: spin foam models, group field theory, simplicial quantum gravity. We have discussed in detail the various conceptual and technical issues to be addressed. 

We focused in particular on collective observables, that is observables depending on large numbers of the microscopic, discrete data defining generic quantum states. This type of observables are indeed the most relevant for the continuum approximation and the extraction of effective physics from quantum gravity at large scales.

Through this type of reasoning we are implementing some notion of coarse graining: normally, the scale $\epsilon$ is
thought to control the degree of refinement at which the geometrical data are treated, representing the UV cutoff, from a Wilsonian perpsective. However,
in this approach in which we move from single edges to extended portions of the graphs, the true scale at which the geometry is sampled depends also on the size of the
objects we are using to encode the geometry. 

When considering such observables, a problem arises: if one uses quantum states that are coherent with respect to microscopic, fundamental observables, and have a factorized structure in terms of states associated to the edges of the same, then quantum fluctuations for collective observables scale with the number of edges. This is true for all the coherent states that have been proposed to date in the quantum gravity literature. This makes them unsuitable for the reconstruction of continuum, classical geometries at large scales. 

We have then identified a general, and rather straightforward, solution to this issue: defining a coherent state based on a different tensorial structure, adapted to the collective classical observable one wishes to approximate. This is indeed what is done, as we recalled, in the much simpler case of many-particle systems in non-relativistic quantum mechanics, where key collective observables are the total momentum of the system and the position of its center of mass. 

We then specialized our general construction to one interesting choice of collective observable: the total electric flux, that is the average of the triad 2-form over a (large) surface embedded in the spatial manifold, and thus generically intersecting many links of any embedded graph. The motivation for choosing this observable was its simplicity, but also the fact that in \cite{OPS} we have defined a new type of coherent states that is especially suitable for the semiclassical approximation of quantum fluxes. We also identified the conjugate variable  to the total flux, at least in some approximation. 

We then defined a collective coherent state adapted to this conjugate pair, which peaks on the classical values and has dispersions saturating the Heisenberg uncertainty, {\it independently of the number of fundamental variables involved in its definition}.  Obviously, the same state would instead give large fluctuations on microscopic observables and it is thus only convenient when considering a continuum approximation at large scales. Another property that we expect our type of states to satisfy, although we have not proven it, is to be more insensitive to graph changing operators, that is to maintain their semiclassicality properties under their action. This is relevant for the issue of dynamics, that in LQG (and spin foams/GFTs) is usually encoded in graph-changing evolution operators. So these states are also expected to be closer to dynamical coherent states: the fluctuations are not going to become too large under evolution, again as far as collective, macroscopic observables are concerned. 

Beside discussing the general construction of these collective observables and of the corresponding coherent state, we also gave an explicit example of it in all its details. This clarifies the various steps of the construction, but also the conceptual subtleties and computational difficulties of the same. 

Finally, we have summarized the steps involved in the construction of generic collective observables and of the corresponding coherent states, as a template for future work.

\

Our results are clearly a first step towards a more complete understanding of the continuum approximation in quantum gravity, at the kinematical level of quantum states, a difficult issue that is being actively investigated, in a dynamical context, for example, in \cite{bianca} from a lattice gauge theory perspective, and in \cite{GFTfluid, lorenzoGFT, vincent, GFThydro, effHamilt, tensor,critical, GFTrenorm} in the context of group field theory and tensor models, not to mention the extensive work in the context of (causal) dynamical triangulations \cite{cdt}. 

The immediate next steps in the same direction involve for example considering other types of collective observables, that could be useful to characterize a continuum semi-classical geometry (volume or areas of extended regions, average intrinsic curvature, etc), and the corresponding coherent states. 

One would like to define a {\it complete} set of such collective observables, in the sense that requiring coherence properties with respect to all of them (and their conjugate variables) allows to specify completely the coherent state associated to a given graph (that is, including what we called $\psi_{rest}$). 

A more ambitious, but not less pressing issue is to construct coherent states that include a sum over graphs in their definition. This means turning on the statistical features of quantum geometry. It would be interesting to analyze whether these statistical features helps to achieve a better approximation of classical collective quantities, forces to modify the choice of quantum state associated to each graph (that is, making the choice of a graph-based coherent state inconvenient) or helps to solve any of the many issues involved in the construction of proper coherent states for quantum gravity \cite{stw}. 

On a more physical level, one would like to use collective coherent states of the type we constructed, capturing the dependence on a few key macroscopic observables, to study large scale dynamics of geometry, in particular in a cosmological context. For example, they could help in understanding the relation between a more fundamental quantum gravity theory and (loop) quantum cosmology \cite{lqc}, when the variables selected are the (average) scale factor (or volume) and its conjugate. 

Of course, this is a difficult task, because it involves upgrading our construction to dynamical coherent states and/or define some effective dynamics for the mean values of simple collective observables. A context where this could be tried, in addition to canonical quantum gravity, is group field theory \cite{GFT}, as a field theory formalism is the best suited for the study of collective dynamics. Doing this would also involve the difficult problem of studying the impact of fluctuations on the mean field dynamics, that is tackling the problem of geometric averaging at the quantum level. 

Another avenue to be investigated is the coupling of gravity with other degrees of freedom, following the lines of \cite{gambinipullin} and \cite{ST} to compute dispersion relations on a spacetime described by the coherent state for the gravity sector. In doing this, the sprinkling procedure, as described in the last section, should be fully understood.

Finally, it would be interesting to introduce perturbations to the classical metric. This would require the construction of the $\psi_{rest}$ part of the wave function and would allow us to study the action of 3d-diffeos on the state. One is interested here in understanding the diffeo invariant degrees of freedom, as this is fully understood for perturbations in the classical theory.

\subsection*{Acknowledgements}

This work has been funded by a Sofja Kovalevskaja Award of the A. von Humboldt Foundation, which is gratefully acknowledged. We thank F.T. Falciano, S.E. Jor\'as, N. Pinto Neto and I. Pr\'emont-Schwarz for useful discussions.

\end{document}